\documentclass{article}
\usepackage{spconf,amsmath,graphicx}
\usepackage[noend]{algpseudocode}
\usepackage{amssymb}
\usepackage{hyperref}
\usepackage{rotating}
\usepackage{multirow}
\usepackage{todonotes}
\usepackage[export]{adjustbox}
\usepackage{algorithm,algorithmicx}
\ninept

\title{Joint learning of cartesian undersampling and reconstruction for accelerated MRI}

\name{Tomer Weiss$^{\star}$, Sanketh Vedula$^{\star}$, Ortal Senouf$^{\star}$, Oleg Michailovich$^{\dagger}$, Michael Zibulevsky$^{\star}$, Alex Bronstein$^{\star}$}

			\address{$^{\star}$ Technion, Israel.
			    $^{\dagger}$ University of Waterloo, Canada.}

\begin{document}
%
\maketitle



%


\maketitle

\begin{abstract}
Magnetic Resonance Imaging (MRI) is considered today the golden-standard modality for soft tissues. The long acquisition times, however, make it more prone to motion artifacts as well as contribute to the relative high costs of this examination. Over the years, multiple studies concentrated on designing reduced measurement schemes and image reconstruction schemes for MRI, however these problems have been so far addressed separately. On the other hand, recent works in optical computational imaging have demonstrated growing success of simultaneous learning-based design of the acquisition and reconstruction schemes manifesting significant improvement in the reconstruction quality with a constrained time budget. Inspired by these successes, in this work, we propose to learn accelerated MR acquisition schemes (in the form of Cartesian trajectories) jointly with the image reconstruction operator. To this end, we propose an algorithm for training the combined acquisition-reconstruction pipeline end-to-end in a differentiable way. We demonstrate the significance of using the learned Cartesian trajectories at different speed up rates.\\
Code available at \href{https://github.com/tomer196/fastMRI-Cartesian}{\textcolor{blue}{https://github.com/tomer196/fastMRI-Cartesian}}.
\end{abstract}


\section{Introduction}
\vspace{-0.2cm}
Magnetic Resonance Imaging (MRI) is a leading modality in medical imaging due to its non-invasiveness, lack of harmful radiation, and excellent contrast and resolution. However, its relatively long image acquisition time currently complicating the use of MRI in many applications such as dynamic imaging and in emergency rooms. During the past few years, compressed sensing \cite{lustig2007sparse} and later deep learning \cite{hammernik2018learning,sun2016deep,zbontar2018fastmri} have been in the forefront of MR image reconstruction, leading to great improvement in image quality with reduced scan times. 
Most studies applying deep learning to improve accelerated MR imaging have concentrated on the reconstruction stage, trying to restore a high quality image from a reduced set of measurements obtained by sub-sampling the $k$-space (i.e., the Fourier domain in which the image is directly acquired). A recent line of works suggested to also learn the acceleration, or the $k$-space sub-sampling scheme. In \cite{gozcu2018learning}, the authors proposed to learn a sampling scheme optimized for off-the-shelf fixed reconstruction methods, which does not fully exploit the strengths of simultaneously learning sampling and reconstruction. In \cite{2019arXiv190101960D}, the authors learned an arbitrary $k$-space sub-sampling together with the reconstruction. However, the decimation rate controlling the speedup factor is not imposed within the pipeline but introduced as an additional term in the loss function. The major drawbacks of this method is the lack of control of the number of measurements, and unrealistic learned trajectories that might be impractical to implement in a real MRI machine or result in lower speedup than the theoretically computed one.

In \cite{zhang2019reducing} and \cite{jin2019self}, the authors propose an active acquisition method using two neural network models: one for the reconstruction, and another one for selecting the next line to be acquired in the $k$-space (the lines are restricted to a Cartesian grid). In each cycle, the current reconstructed image and the $k$-space are used as an input to the model, selecting the next sample. The former paper relies on an uncertainty map-based mechanism, whereas the latter uses a Monte-Carlo tree search. The major drawback of these methods are their complexity, which is especially acute due to the stringent low-latency requirement of the real-time acquisition setting. 

{\bf Contributions.} In this paper, we propose to simultaneously train a differentiable forward model ($k$-space sub-sampling) and its inverse (the image reconstruction pipeline). The training is performed end-to-end without imposing any complex line selection mechanism. We show that by implementing this straightforward trainable sampling-reconstruction scheme we achieve similar improvement margins compared to more complicated schemes.
\vspace{-0.2cm}

\section{Method}
\vspace{-0.2cm}
\begin{figure*}[!ht]
   \centering
\includegraphics[width=15cm]{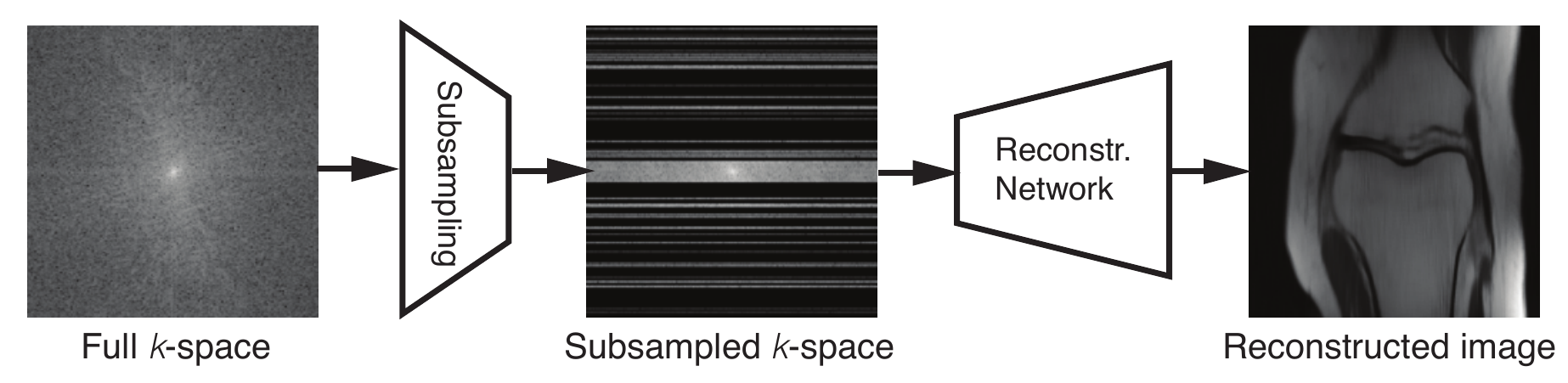}.
    \caption{Accelerated MRI acquisition and reconstruction. In standard approaches, the sub-sampling mask is fixed, and the reconstruction network is trained to obtain the highest quality image. In our approach, the sub-sampling is a trainable layer in an end-to-end neural network comprising both sensing and reconstruction. }
    \label{fig:pipline} 
\vspace{-0.2cm}
\end{figure*}

In our pipeline, we perform the sub-sampling in the $k$-space domain and the reconstruction in the image domain, following an inverse Fourier transform, as described schematically in Fig. \ref{fig:pipline}. In what follows, we describe each of these two ingredients in greater detail.
\vspace{-0.2cm}
\subsection{Sub-sampling layer}
As depicted in Fig. \ref{fig:pipline}, the sub-sampling layer receives a fully sampled $k$-space, denoted as $x$ and outputs the sub-sampled version $y=\Phi x$, where $\Phi$ is a binary sub-sampling mask. Being restricted to Cartesian trajectories, the sub-sampling mask $\Phi$ is a column vector, the length of the number of rows, $M$, of the k-space matrix.  

 In order to allow end-to-end joint training of the sub-sampling mask and the reconstruction, the binary nature of the sub-sampling operation must be taken into account.
We follow the methodology of \cite{2016arXiv160703343I,2015arXiv151100363C} proposing to keep two versions of the mask: binary, denoted as $\Phi$, and continuous, denoted as $\Phi_c$. The two versions are used as follows:
\begin{enumerate}
  \item During forward and back propagation for calculating the gradients, the binary version $\Phi$ is used; 
  \item The gradient step with the calculated weight update $\delta \Phi$ is applied to the continuous $\Phi_c$;
  \item The continuous mask $\Phi_c$ is binarized as follows to produce an updated version of  $\Phi$:
  $$
 (\Phi)_{ij} = \begin{cases}
             1,  & \text{if } (\Phi_c)_{ij} \ge \tau,  \\
             0,  &   \text{otherwise}\\
       \end{cases} \quad
  $$
where $\tau$ is determined as the upper $q$-tile of the values of $\Phi_c$ with $q$ denoting the decimation rate.
\end{enumerate}

\noindent In order to initialize the sub-sampling layer, we first generate a semi-random mask, by choosing a pre-determined amount of low-frequency rows, followed by randomly selecting additional higher frequency rows until reaching the desired decimation rate. Consequently, the continuous mask $\Phi_c$ is initialized by assigning a random value from the uniform distribution $U\big(0.5,1\big)$ to each row selected in $\Phi$, and a random value from the uniform distribution $U\big(0,0.5\big)$ otherwise.

\vspace{-0.2cm}
\subsection{Reconstruction network}
\vspace{-0.1cm}
As the inverse model, we used a multi-resolution encoder-decoder network with symmetric skip connections, also known as the U-net architecture \cite{ronneberger2015u}. U-net is widely-used in medical imaging reconstruction tasks in general and in MRI reconstruction in particular \cite{zbontar2018fastmri}. The network is henceforth denoted as $G_\theta$, with $\theta$ representing its learnable parameters. The input to the network is the distorted MR image, generated by the inverse Fourier transform of the sub-sampled $k$-space, $z_{in}  =\mathcal{F}^{-1}(\Phi x)$; the output is the reconstructed fully sampled MR image $\hat{z}=G_\theta(z_{in})$ estimating the groundtruth image $z = \mathcal{F}^{-1}(x)$.\\ \\The full flow of training our sampling-reconstruction scheme is summarized in Algorithm \ref{Algo:LearnedFastMRI}.

	\begin{algorithm}
		\caption{Learning Fast MRI}\label{Algo:LearnedFastMRI}
	\begin{algorithmic}[1]
		\Statex{\textbf{input: } $x$ - Full MRI k-space}
		\State {\textbf{Initialize $\Phi$:} random Cartesian binary sampling mask, complying with the speedup factor $q$}
		\State{\textbf{Initialize $\Phi_c$:} according to the binary values, with random values from a uniform distribution}
		\While{(not converged)}
		    \State {\textbf{Apply binary mask} $y=\Phi x$ to obtain a reduced set of measurements $y$}
		    \State{\textbf{Forward pass:} calculate loss  $L\big(\mathcal{F}^{-1}(x),G_\theta(\mathcal{F}^{-1}(\Phi x))\big)$}
		    \State{\textbf{Backward pass:} calculate gradients $\delta \Phi = \nabla_\Phi L$, $\delta \theta = \nabla_\theta L$}
		    \State{\textbf{Update weights} $\Phi_c$, $\theta$ ($\Phi_c$ is updated using $\nabla\Phi$)}
		    \State{\textbf{Update binary mask} $\Phi$ by binarizing $\Phi_c$ }
 		\EndWhile
 	\end{algorithmic}
 	\end{algorithm}
\vspace{-0.3cm}
\section{Experiments and discussion}
\vspace{-0.2cm}
\subsection{Dataset}
Data used in the preparation of this article were obtained from the NYU fastMRI Initiative database (\url{fastmri.med.nyu.edu}) \cite{zbontar2018fastmri}. The NYU fastMRI investigators only provided the data but did not participate in the analysis or in the writing of this report\footnote{More details regarding the dataset acquisition and split can be found in \cite{zbontar2018fastmri}.}.
    The fastMRI dataset contains $1372$ knee MRI volumes. Since the core of our work is learning the $k$-space sub-sampling scheme and the provided test set contains already sub-sampled MR images, we have used the training set for our experiments and divided it into two sets: one containing $973$ volumes ($34700$ slices) for training and validation, and the other one containing $199$ volumes ($7100$ slices) for testing. 
\vspace{-0.3cm}
\subsection{Settings}
\vspace{-0.2cm}
\label{sec:settings}
\textbf{Training. } We trained the U-net with the RMSprop \cite{tieleman2012lecture} solver, while the sub-sampling layer was trained with simple stochastic gradient with momentum \cite{momentum99}, both with learning rate of $0.001$. The loss function was set to the $L_1$ error. For all our experiments, we work with 2D slices processed using 2D CNNs.
It should be mentioned that as the scope of this work was to show the benefit of simultaneously optimizing the sub-sampling pattern with the reconstruction model, we do not perform any architectural search for the reconstruction network, and use the U-net configuration that has been used in \cite{zbontar2018fastmri}).\\
\textbf{Different decimation rates.} We performed our experiment with different decimation rates (speedup factors) of the $k$-space: $4$, $8$, $12$, and  $16$, corresponding to $q=25\%$, $12.5\%$, $8.3\%$ and $6.25\%$ of the measurements.\\
\textbf{Different central fraction sizes.} In this experiment, we fixed the decimation rate and performed the evaluation while initializing the sub-sampling mask with different sizes of the central fraction of the $k$-space, that is, varying the rate of the active low frequency rows in the $k$-space. For example, at decimation rate of $4$ ($25\%$ of the measurements) we initialize with the central fraction between $0\%$ to $25\%$.
In all experiments we compared our method to a fixed mask scenario, where only the reconstruction model was trained. These fixed masks served as the initialization for the learned mask experiments. 
\vspace{-0.3cm}
\subsection{Results and discussion}
\begin{figure*}[!t]
\begin{tabular}{c c}
\includegraphics[width=0.3\textwidth,height=0.2\textwidth]{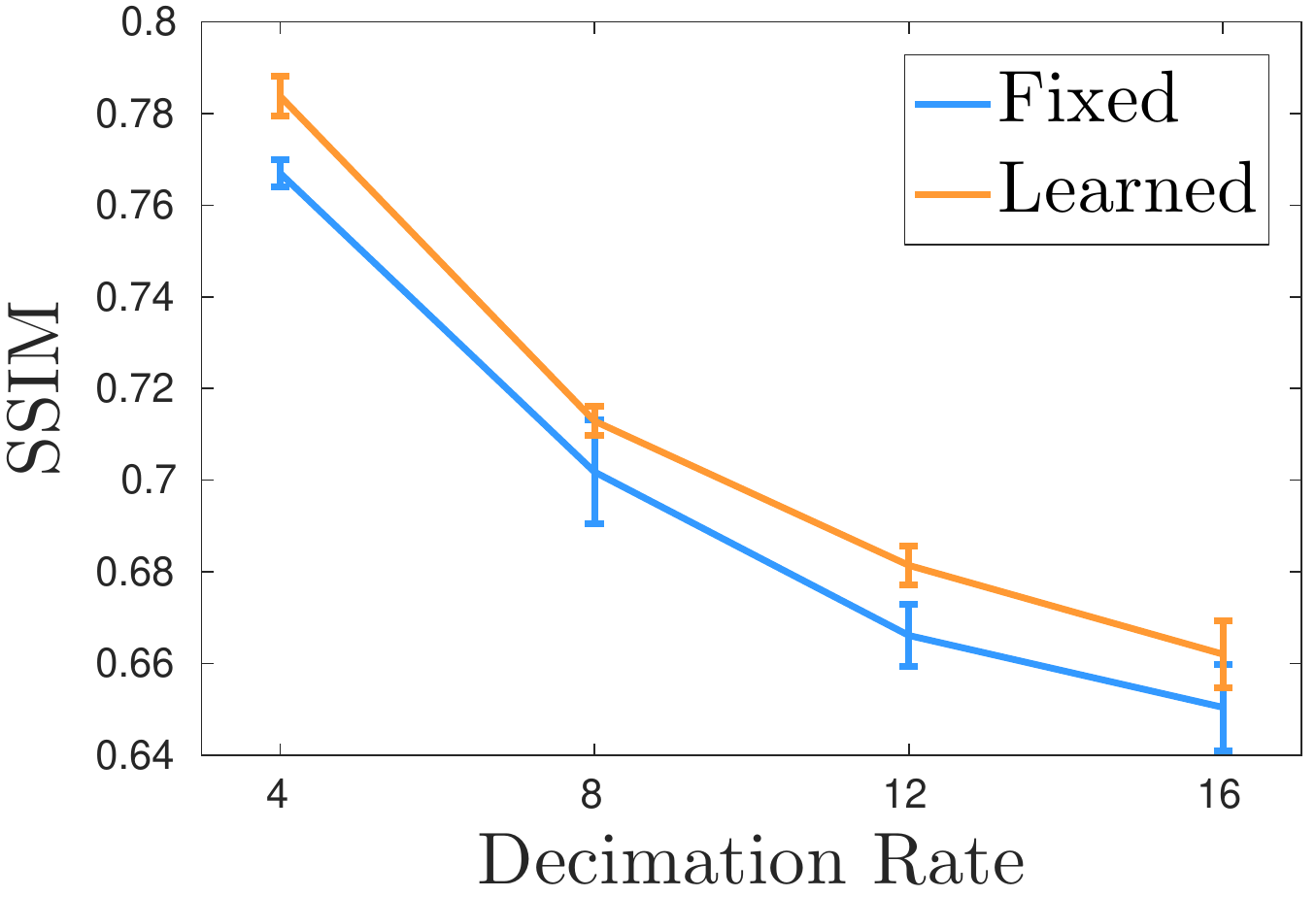}&
\includegraphics[width=0.3\textwidth]{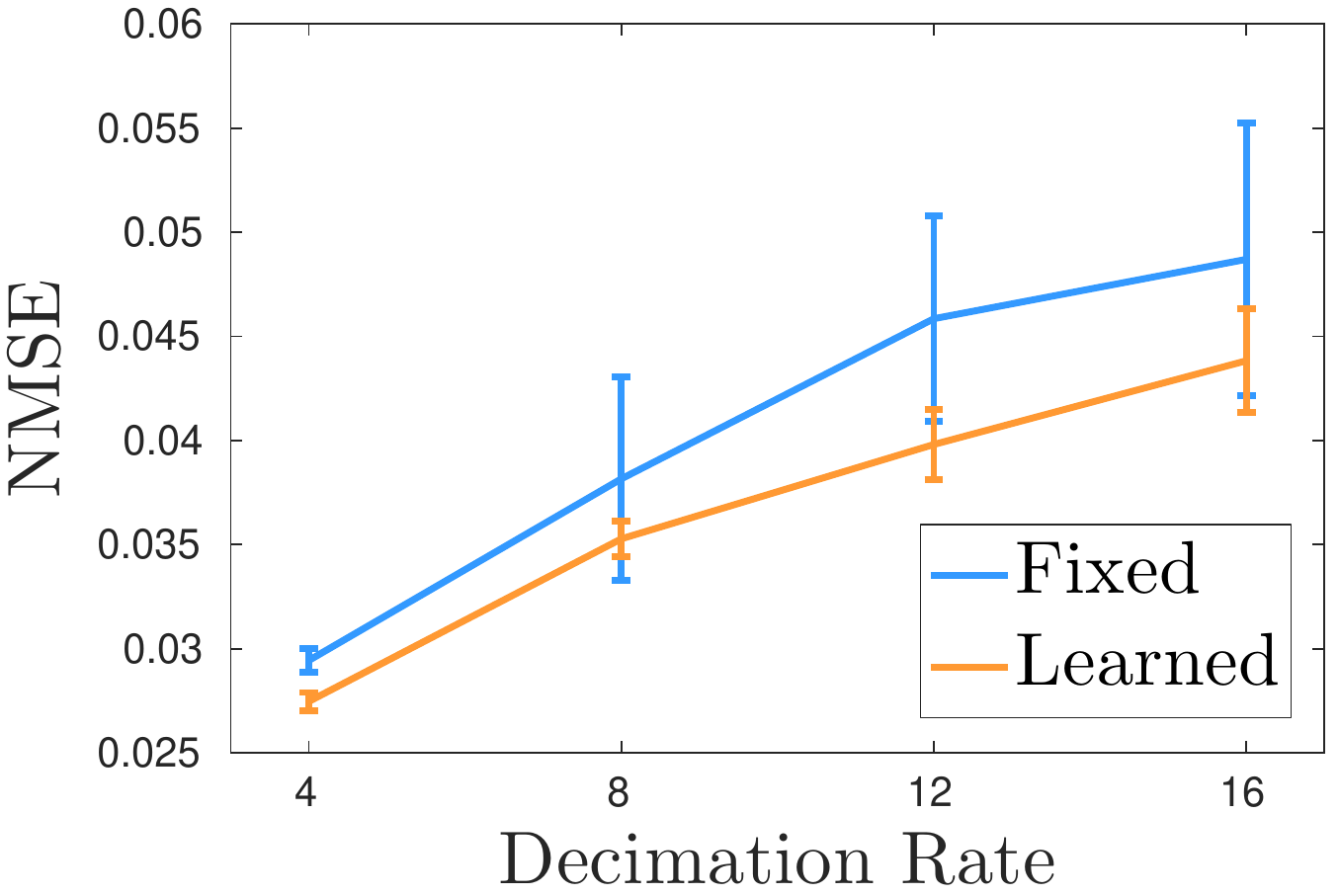}
\includegraphics[width=0.3\textwidth,height=0.2\textwidth]{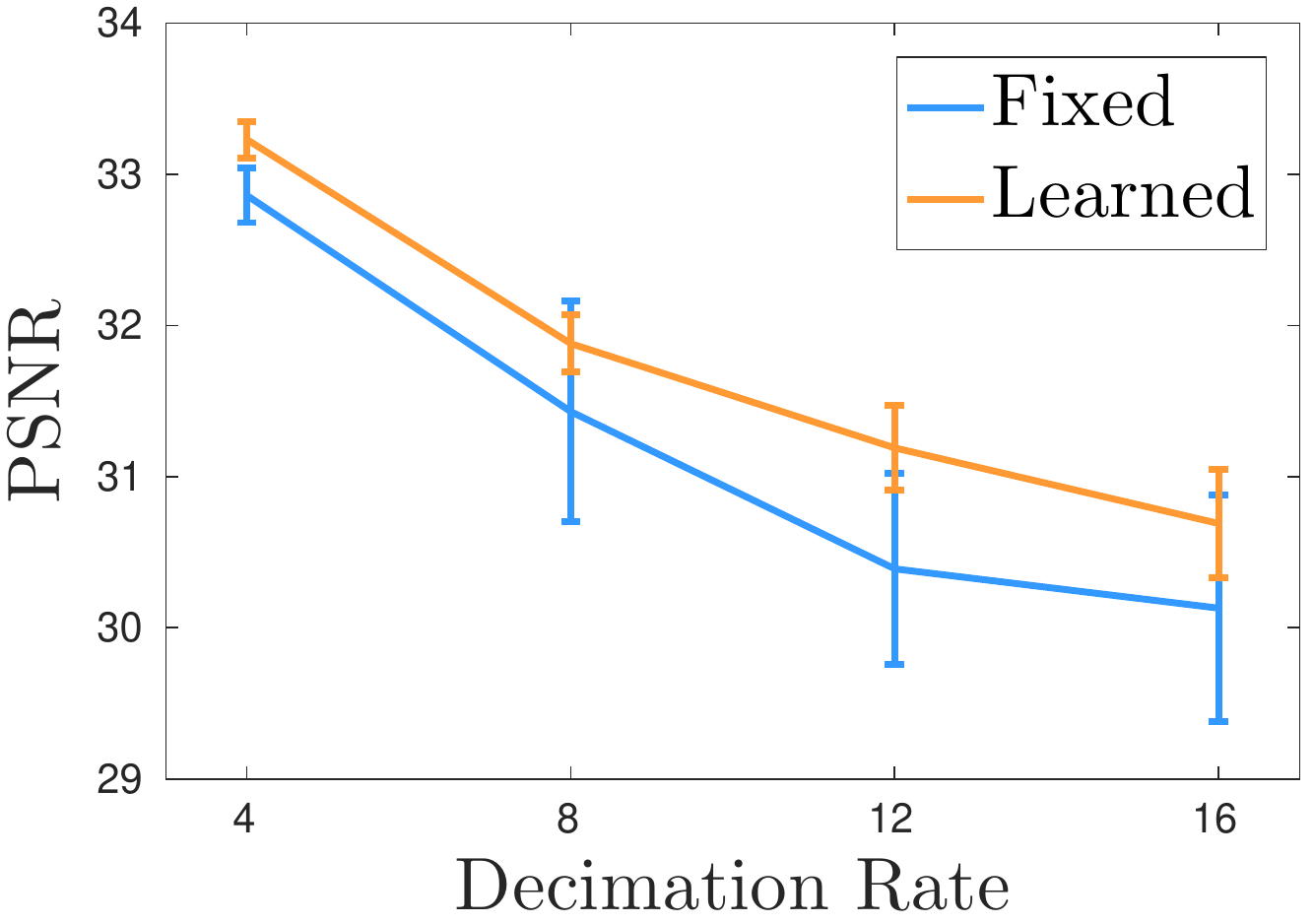}
\end{tabular}
\caption{Image reconstruction quality using fixed and learned masks for different decimation rates. Plotted are the average PSNR, SSIM \& NMSE and standard deviation across different initializations.}
\label{fig:plot} 
\vspace{-0.3cm}
\end{figure*}

\begin{figure}[!b]
   \centering
\includegraphics[width=0.38\textwidth]{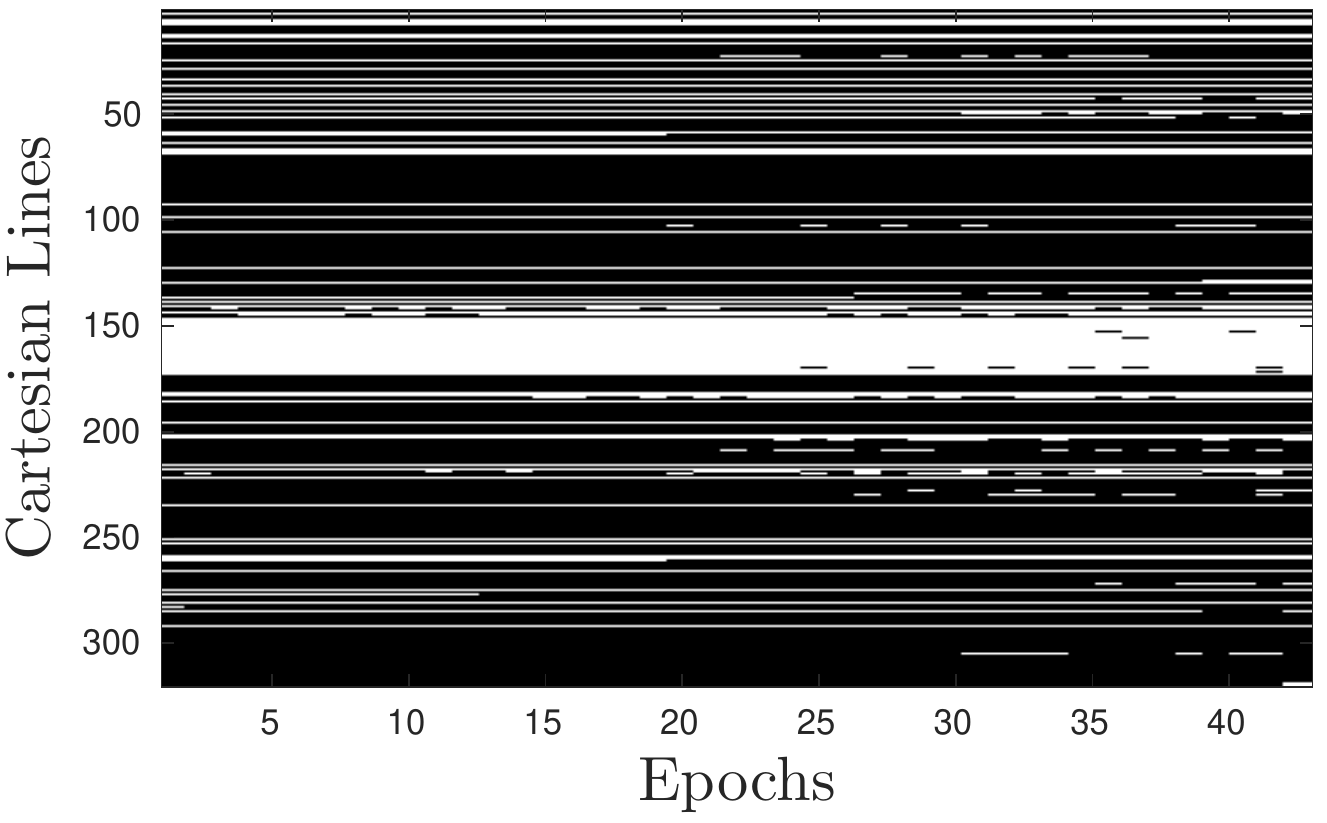}\\
\includegraphics[width=0.38\textwidth]{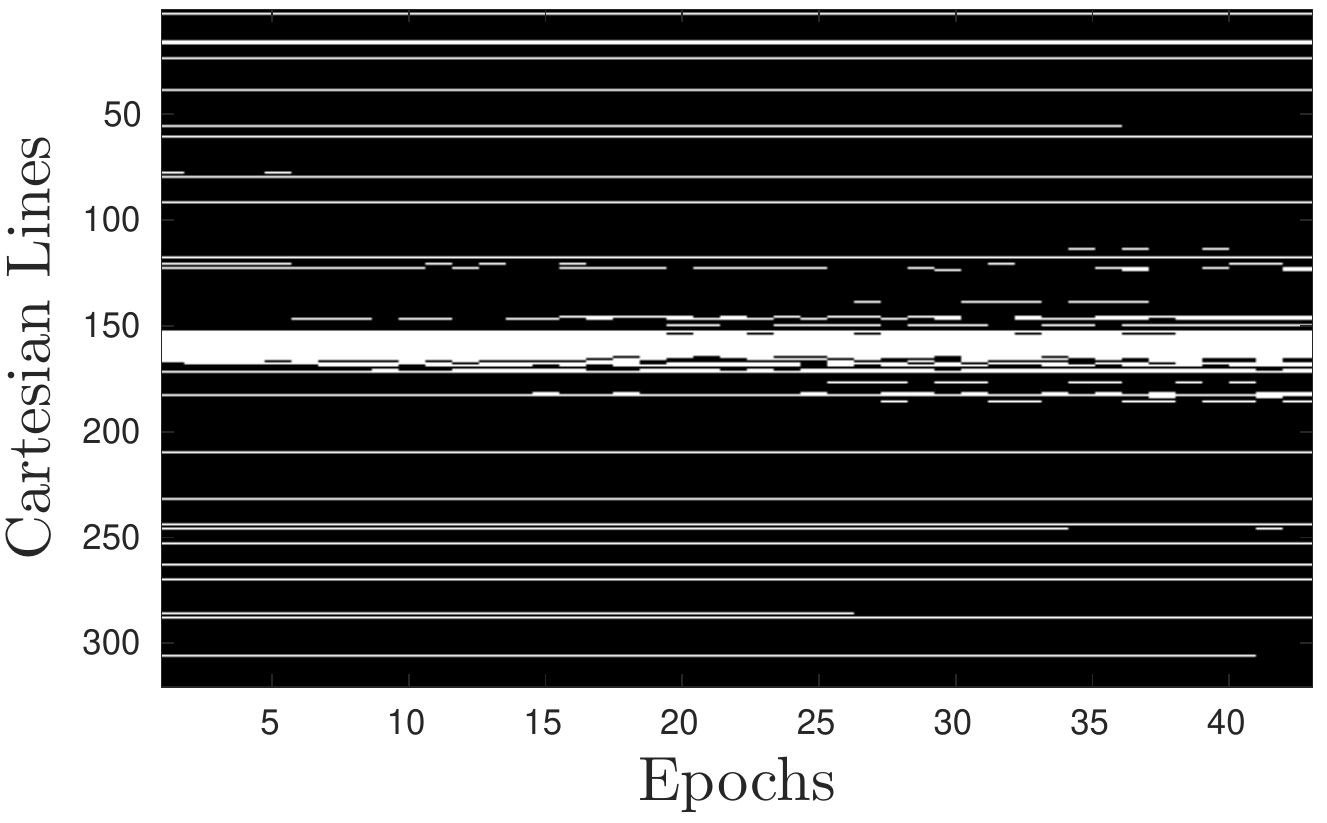}\\
\caption{Mask evolution during training. Training progress is shown on the horizontal axis in epochs. 4-fold acceleration (top) and 8-fold acceleration (bottom).}    \label{fig:mask_evaluation} 
    \vspace{-0.3cm}
\end{figure}

\begin{figure*}[!hb]
   \centering
      \addtolength{\tabcolsep}{-6.5pt}

\begin{tabular}{c c c c c c}
Ground truth&4&8&12&16&\\
\includegraphics[width=0.2\textwidth]{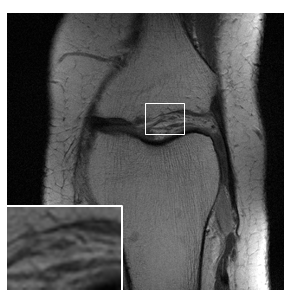}&
\includegraphics[width=0.2\textwidth]{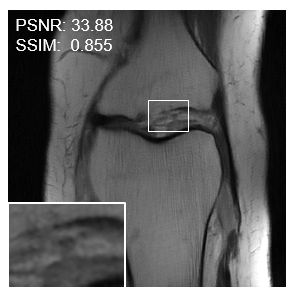}&
\includegraphics[width=0.2\textwidth]{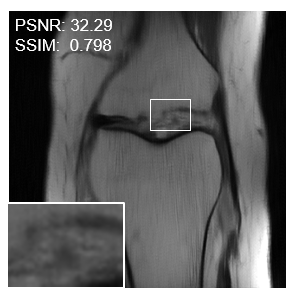}&
\includegraphics[width=0.2\textwidth]{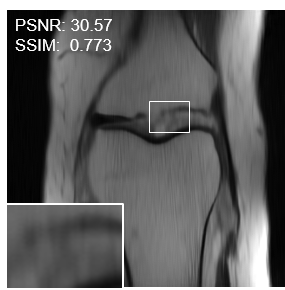}&
\includegraphics[width=0.2\textwidth]{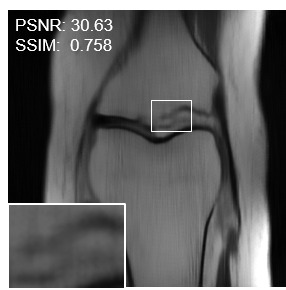}&
\rotatebox{90}{\hspace{0.8cm} Fixed mask}\\
\vspace{-9pt}
&
\includegraphics[width=0.2\textwidth]{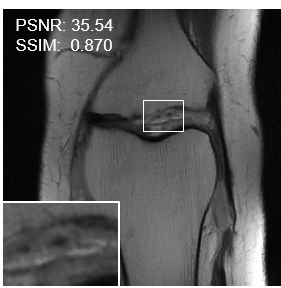}&
\includegraphics[width=0.2\textwidth]{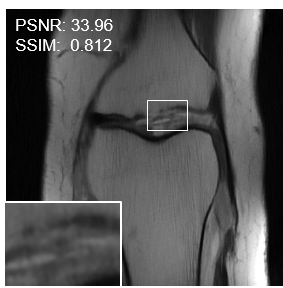}&
\includegraphics[width=0.2\textwidth]{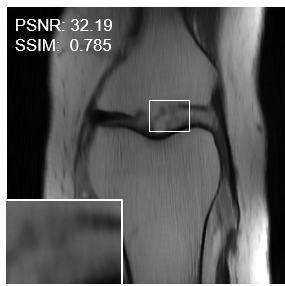}&
\includegraphics[width=0.2\textwidth]{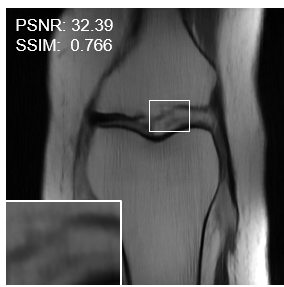}&
\rotatebox{90}{\hspace{0.8cm} Learned mask}\\
&
\includegraphics[width=0.2\textwidth]{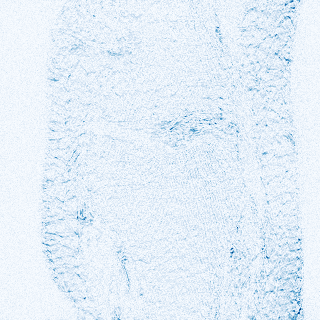}&
\includegraphics[width=0.2\textwidth]{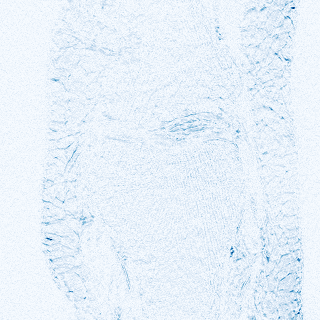}&
\includegraphics[width=0.2\textwidth]{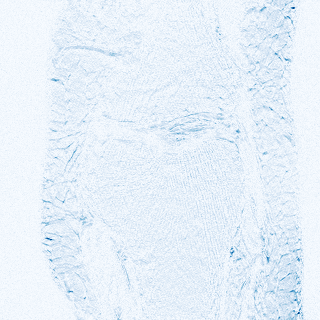}&
\includegraphics[width=0.2\textwidth]{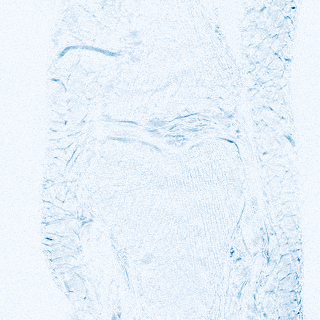}&
\rotatebox{90}{\hspace{0.8cm} Fixed mask}\\
&
\includegraphics[width=0.2\textwidth]{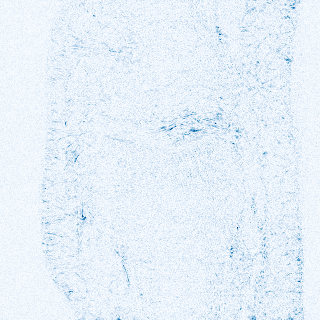}&
\includegraphics[width=0.2\textwidth]{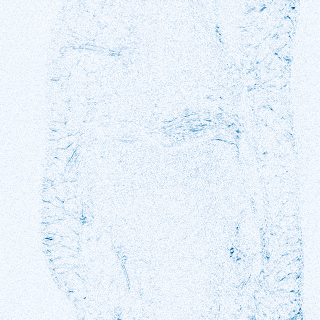}&
\includegraphics[width=0.2\textwidth]{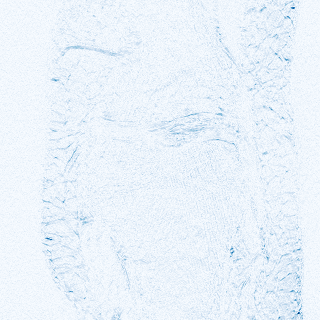}&
\includegraphics[width=0.2\textwidth]{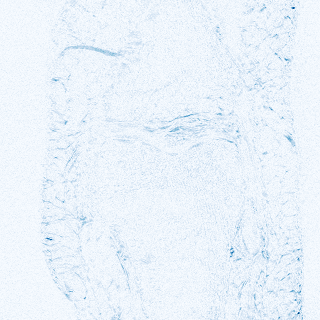}&
\rotatebox{90}{\hspace{0.8cm} Learned mask}\\
&
\includegraphics[width=0.2\textwidth, cfbox=red 0.5pt 1pt]{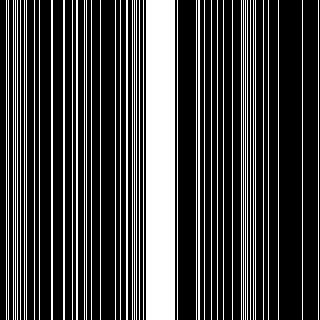}&
\includegraphics[width=0.2\textwidth, cfbox=red 0.5pt 1pt]{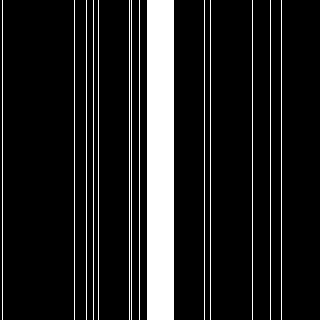}&
\includegraphics[width=0.2\textwidth, cfbox=red 0.5pt 1pt]{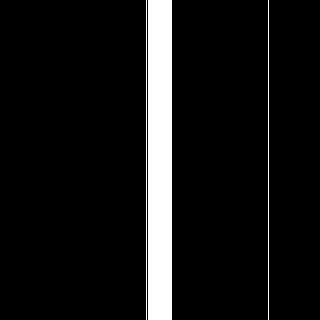}&
\includegraphics[width=0.2\textwidth, cfbox=red 0.5pt 1pt]{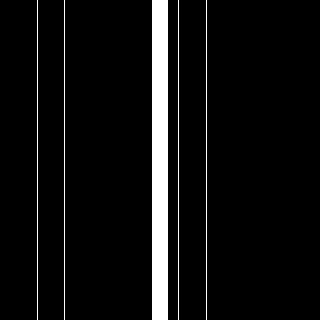}&
\rotatebox{90}{\hspace{0.8cm} Fixed mask}\\
&
\includegraphics[width=0.2\textwidth, cfbox=blue 0.5pt 1pt]{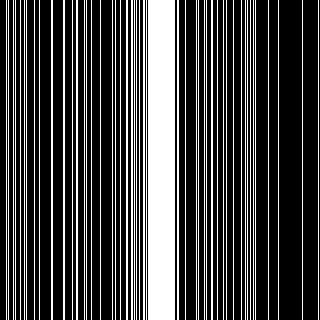}&
\includegraphics[width=0.2\textwidth, cfbox=blue 0.5pt 1pt]{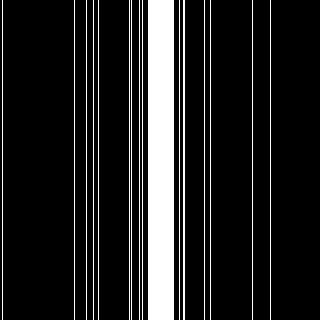}&
\includegraphics[width=0.2\textwidth, cfbox=blue 0.5pt 1pt]{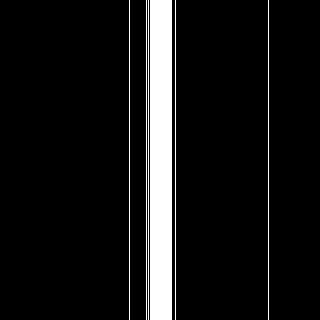}&
\includegraphics[width=0.2\textwidth, cfbox=blue 0.5pt 1pt]{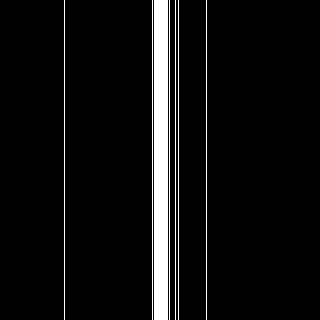}&
\rotatebox{90}{\hspace{0.8cm} Learned mask}\\
\end{tabular}
   \addtolength{\tabcolsep}{1pt}

    \caption{Visual comparison between fixed/learned masks for different decimation rates. Rows 1-2 show the reconstructed image generated by the Fixed/Learned mask models; rows 3-4 depict the difference map between the reconstruction and the ground truth (darker represents higher error); rows 5-6 show the fixed initial mask and the learned mask. The masks are rotated $90^\circ$ counterclockwise for the ease of comparison.}
    \label{fig:Visual Results} 
\vspace{-0.3cm}
\end{figure*}

For the quantitative evaluation of our method, we selected the commonly used peak-signal-to-noise-ratio (PSNR), and structural-similarity measure (SSIM), and the normalized mean-squared-error (NMSE), portraying both pixel-to-pixel and perceptual similarity. Fig. \ref{fig:plot} depicts the image distortion (in terms of PSNR, SSIM \& NMSE) as a function of the decimation rate; standard deviation was calculated on repeated experiments with different initial central fraction sizes (as previously explained in Section \ref{sec:settings}). The results suggests that our method outperforms the fixed mask setting across all decimation rates; the improvement is in the range of $0.45-0.8$ dB PSNR, $0.012-0.017$ SSIM points, and $0.002-0.006$ in NMSE. It is interesting to notice the increased improvement in the higher decimation rates, and that the proposed mask training method is less sensitive to initialization -- notice that the standard deviation for the learned sampling is lesser than the fixed alternative (Fig. \ref{fig:plot}).
Visual inspection of one slice, displayed in Fig. \ref{fig:Visual Results}, supports this quantitative evaluation. The zoomed-in area displays better contrast and sharpness with our learned mask, comparing to the fixed masks models. Additionally, observing the scaled difference maps, one can see that our method produced lower errors than the fixed mask reconstructions, especially in the highly detailed areas.      

\textit{{\bf Comparison to the state-of-the-art.}} A fair quantitative comparison to the recently proposed learned accelerated MRI methods is not practical for several reasons, including the lack of uniform evaluation metrics and test sets, and the lack of availability of the competing pre-trained models. However, we can still discuss the improvement margins between the fixed and learned mask experiments for all methods. In \cite{gozcu2018learning}, the authors reported margins of about $2.5$ dB PSNR, but their method was tested on a very limited test set ($60$ slices compared to $7100$ in our experiments). In \cite{2019arXiv190101960D}, arbitrary (non-Cartesian)  masks were learned, and by that the sub-sampling rates were not indicative of the acceleration rates, therefore it is not comparable with our learned Cartesian masks. In \cite{jin2019self}, the authors reported  margins similar to ours (about $0.5dB-1$ dB PSNR) on both knee and cardiac data sets with a $4$-fold acceleration factor; no results were reported for other acceleration rates. In \cite{zhang2019reducing}, the authors reported similar results to ours (improvement by about $0.02$ SSIM points); no PSNR values were reported. It is important to emphasize that while exhibiting similar improvement on similar experiments, the last two methods suggest a much more complex processing pipeline.

\textit{Mask evolution.}
Fig. \ref{fig:mask_evaluation} displays the evolution of the the mask during training for both $4-$ and $8-$fold acceleration rates. A key observation is that the model "selects" different frequency lines than the ones of the initial mask, but still preserves similarity to it. This implies that while the final mask is depended on the initialization, the learning process consistently improves the performance of the mask.

\vspace{-0.3cm}

\section{Conclusion and future directions}

We have demonstrated, as a proof-of-concept, that learning simultaneously the sub-sampling pattern and the reconstruction network improves the end image quality of an MR imaging system. The method can simply be implemented on the current scanners using the Cartesian sampling in a multi-shot setting.
It should be mentioned that since the results were affected by the mask initialization, we can assume the models have not reached the globally optimal configuration -- otherwise, all patterns would have converged to similar performance. This calls for better optimization techniques that are more robust to initialization -- a statement that is true for practically every deep learning model.

The main limitation of our work is the restriction of the optimal sub-sampling patterns to Cartesian trajectories, which are the most commonly used schemes in MR machines today. However, higher speed-ups can be gained by going beyond the Cartesian sampling, while still adhering to the constraints of the machine. Research has been conducted on designing optimal non-Cartesian trajectories (see \cite{sparkling2019Lazarus} and references therein). In a recent work \cite{pilot2019weiss}, we show that by learning arbitrary optimal $k$-space trajectories and the reconstruction model simultaneously while imposing the machine physical constraints, we achieve a greater quality of the resulted images comparing to Cartesian acquisition, while maintaining the trajectories physically feasible.  In the future, we plan to conduct prospective accelerated acquisitions on real MRI scanners using these learned protocols.

\vfill\pagebreak
\clearpage

\bibliographystyle{IEEEbib}
\bibliography{main}

\end{document}